\input epsf.tex
 \documentstyle[12pt]{article}

\newcommand{\bmat}{\left(\begin{array}}
\newcommand{\emat}{\end{array}\right)}
\def\NPB{Nucl. Phys. B}

\def\yzero{\smash{\hbox{$y\kern-4pt\raise1pt\hbox{${}^\circ$}$}}}

\def\b{\beta}

\def\beq{\begin{equation}}
\def\eeq{\end{equation}}
\def\beqa{\begin{eqnarray}}
\def\eeqa{\end{eqnarray}}

\def\-{\hphantom{-}}

\def\s2{\frac{1}{\sqrt2}}

\def\beq{\begin{equation}}
\def\eeq{\end{equation}}
\def\beqa{\begin{eqnarray}}
\def\eeqa{\end{eqnarray}}

\def\IF{\relax{\rm I\kern-.18em F}}
\def\II{\relax{\rm I\kern-.18em I}}
\def\IP{\relax{\rm I\kern-.18em P}}
\def\IC{\relax\hbox{\kern.25em$\inbar\kern-.3em{\rm C}$}}
\def\IR{\relax{\rm I\kern-.18em R}}

\def\cp{{\cal P}}

\def\Dsl{\,\raise.15ex\hbox{/}\mkern-13.5mu D} 
\def\IZ{Z\kern-.4em  Z}

 \def\cp#1{\relax\ifmmode {\IP\kern-2pt{}_{#1}}\else $\IP\kern-2pt{}_{#1}$\=fi}

\catcode`\@=11
\newdimen\@rotdimen
\newbox\@rotbox

\def\@vspec#1{\special{ps:#1}}
\def\@rotstart#1{\@vspec{gsave currentpoint currentpoint translate
   #1 neg exch neg exch translate}}
\def\@rotfinish{\@vspec{currentpoint grestore moveto}}
\def\@rotr#1{\@rotdimen=\ht#1\advance\@rotdimen by\dp#1%
   \hbox to\@rotdimen{\hskip\ht#1\vbox to\wd#1{\@rotstart{90 rotate}%
   \box#1\vss}\hss}\@rotfinish}
\def\@rotl#1{\@rotdimen=\ht#1\advance\@rotdimen by\dp#1%
   \hbox to\@rotdimen{\vbox to\wd#1{\vskip\wd#1\@rotstart{270 rotate}%
   \box#1\vss}\hss}\@rotfinish}%
\def\@rotu#1{\@rotdimen=\ht#1\advance\@rotdimen by\dp#1%
   \hbox to\wd#1{\hskip\wd#1\vbox to\@rotdimen{\vskip\@rotdimen
   \@rotstart{-1 dup scale}\box#1\vss}\hss}\@rotfinish}%
\def\@rotf#1{\hbox to\wd#1{\hskip\wd#1\@rotstart{-1 1 scale}%
   \box#1\hss}\@rotfinish}%
\def\rotate{\@ifnextchar[{\@rotate}{\@rotate[l]}}
\def\@rotate[#1]#2{\setbox\@rotbox=\hbox{#2}\@nameuse{@rot#1}\@rotbox}

\catcode`\@=12

\begin{document}

\makeatletter
\@addtoreset{equation}{section} \makeatother
\renewcommand{\theequation}{\thesection.\arabic{equation}}
\pagestyle{empty}
\pagestyle{empty}
\rightline{FTUAM-02-17}
\rightline{IFT-UAM/CSIC-02-20}
\rightline{\today}
\vspace{0.5cm}
\setcounter{footnote}{0}

\begin{center}
{\LARGE{\bf Exact Standard Model Compactifications
from Intersecting Branes}}
\\[7mm]
{\Large{{  Christos ~Kokorelis} }
\\[2mm]}
\small{ Dep/to de F\'\i sica Te\'orica C-XI and 
Instituto de F\'\i sica 
Te\'orica C-XVI}
,\\[-0.3em]
{ Universidad Aut\'onoma de Madrid, Cantoblanco, 28049, Madrid, Spain}
\end{center}
\vspace{3mm}


\begin{center}
{\small \bf ABSTRACT}
\end{center}
\begin{center}
\begin{minipage}[h]{14.5cm}
 We construct six stack D6-brane vacua (non-supersymmetric) that have 
at low energy exactly
the
standard model (with right handed neutrinos).
The construction is based on 
D6-branes
intersecting at angles in $D = 4$ type toroidal orientifolds of type
I strings.
Three $U(1)$'s become massive through their couplings to RR fields
and from the three surviving massless $U(1)$'s at low energies, one is the 
standard model 
hypercharge generator. The two extra massless $U(1)$'s get broken, as
suggested
recently (hep-th/0205147),  by requiring some intersections 
to respect $N=1$ supersymmetry thus supporting the appearance of massless
charged singlets.
Proton and lepton number are gauged symmetries and their 
anomalies are cancelled through a generalized Green-Schwarz
mechanism that gives masses to the corresponding gauge 
bosons through couplings to RR fields. Thus proton is stable and 
neutrinos are of Dirac type with small masses
as a result of a PQ like-symmetry.
The models predict the existence
of only two supersymmetric
particles, superpartners
of $\nu_R$'s.

\end{minipage}                 
\end{center}

\newpage
\setcounter{page}{1}
\pagestyle{plain}
\renewcommand{\thefootnote}{\arabic{footnote}}
\setcounter{footnote}{0}

\section{Introduction}

Obtaining chiral string constructions where the low energy particle content
may be the observable chiral spectrum of the standard model (SM) and
gauge interactions,
is one of the important directions of string theory research.
In this respect semirealistic models have been pursued
along those directions, both in the context
of 4D $N=1$ heterotic compactifications and in orientifold
constructions \cite{see}.

In this work, we will examine
standard model compactifications in the context
of recent constructions \cite{inter1, inter2}, which use
intersecting branes and give 4D non-SUSY models.
So why we have to resort to non-supersymmetric models in our search for
realistic string models ?
In $N=1$ heterotic orbifold compactifications (HOC) the semirealistic
models derived
were supersymmetric (SUSY) and included at low energy
the
MSSM particle content, together with a variety of exotic matter and/or
gauge group factors.
However,
in order to reconcile the observed discrepancy
between the unification of gauge couplings in the MSSM at
$10^{16}$ GeV \cite{ellisnano} and the string scale at 
HOC which is of order $10^{17}-10^{18}$ GeV,
it was assumed that the observed difference should be attributed to the
presence of the
string one loop corrections to the $N=1$ 
gauge coupling constants \cite{kokos1}.
Thus even though it was not
possible for a single model to be found which realized at low energy only
the MSSM it was assumed that this will become possible 
after an extensive study of the different compactification
vacua was performed.
However, the goal of obtaining a particular string compactification
with only the MSSM content was not realized. 
On the other hand in type I compactifications (IC) the string scale is a
free parameter. In addition, recent results suggest
that it is possible in IC to lower the string scale, by having some
compact directions transverse to all stacks of branes, in the TeV region
even without SUSY \cite{anto}. Thus non-SUSY models with a
string scale in the TeV region provide us with a viable alternative
to SUSY models.

For models based on intersecting
branes the main picture involves
localization of fermions
in the intersections between branes \cite{ber}.
In these constructions the
introduction of a quantized NS-NS B 
field \cite{flux} effectively produces
semirealistic models with 
three generations \cite{inter2}. Under T-duality these
backgrounds transform to models with
magnetic deformations \cite{carlo1, carlo2}.

Recently, an interesting class of models was found
that uses four-stacks of
branes and gives exactly the SM content at low energies \cite{luis1}. 
The models were based on
D6-branes intersecting at angles on an orientifolded six-torus
compactification \cite{inter1, inter2}.  
These models have an interesting generalization to classes of
models with
five-stacks of branes \cite{kokos2}. The features of the latter models are
quite similar to those that constitute the main part of this work.
The models
of \cite{luis1, kokos2} share some common
features as proton and
lepton number are gauged symmetries surviving as global
symmetries to low energies, small neutrino masses and a remarkable
Higgs sector
that in
cases is of the MSSM. On the contrary, while the four stack model
is a non-SUSY model and has a variety of sectors with non-SUSY chiral
fields, models that have five-stacks have one additional
unusual feature.
They have some sectors that preserve $N=1$ SUSY, thus a $N=1$
hypermupliplet remains initially massless \footnote{with the scalar
component
receiving eventually a mass}, even though the model is
overall a non-SUSY one. The latter feature, as we will see, is maintained
in the six-stack classes of models presented in this work.

The models of \cite{luis1} have been extended
to describe the first string non-SUSY model examples, of classes of GUT 
structured models, which give exactly the SM at low
energies \cite{kokos3}.
These classes of models \footnote{They are build on a background
of four-stacks of D6 branes, intersecting at angles, 
on an orientifolded $T^6$ background} 
are based on the
Pati-Salam gauge group $SU(4) \otimes SU(2)_L \otimes SU(2)_R$ and represent
at present, the only consistent GUT models in the
context of intersecting branes that are able to produce exactly
the SM at low energies.
These classes of models maintain essential
features of \cite{luis1, kokos2} and in particular the fact that proton is
stable, as the baryon number is an unbroken gauged symmetry, and small
neutrino masses. Also the GUT four-stack classes of
models share the unusual features of the classes of the
five-stack SM's, four-stack
GUT's of \cite{kokos2, kokos3} respectively, that is even
though the models are non-SUSY they do allow for some sectors
to preserve $N=1$ SUSY.
It is quite interesting to note that,
even though it was generally believed that in D6 brane orientifolded six
torus models it was not possible to find an apparent explanation
for lowering the string scale in the TeV region \footnote{since
there were no torus directions transverse to all branes.}, 
in the classes of
GUT models
of \cite{kokos3} this issue was solved. In particular, the models
predict the existence of light weak doublets with mass of order
$\upsilon^2/M_s$, that necessarily needs the string scale to be
less than 650 GeV. 
The latter results are particularly encouraging as
they represent strong predictions in the context of 
intersecting D-brane scenarios and are directly
testable at present or future accelerators.
For additional developments with non-SUSY constructions
in the context
of intersecting branes,
see \cite{allo2, allo3, allo4,
allo5, allo6, allo6aa, allo6a, allo7, allo8}.
For a $N=1$ SUSY construction, in the context of
intersecting branes,
and its phenomenology see \cite{shiu}.

In this work, we will discuss compactifications of intersecting
D6 branes, with exactly the SM at low energies that use six-stacks of
D-branes on an orientifolded six-torus.
Thus we 
practically extending the models of \cite{kokos2} by one more $U(1)$
stack and the models of \cite{luis1} by two more $U(1)$ stacks.
The classes of SM's that we discuss in this work
possess some general features:

\begin{itemize}

\item Even if the classes of models are
non-SUSY overall, they have sectors that preserve $N=1$ supersymmetry.
This unusual feature has never appeared in the context of string
theory before.
It appears that the imposition of SUSY at particular sectors,
of the non-SUSY models, creates the
necessary singlets that break the two extra \footnote{Beyond
the SM gauge symmetries.}
$U(1)$'s symmetries leaving at low energy
exactly the SM.

\item The models predict unexpectedly the existence of only two
SUSY particles, superpartners of right handed neutrinos.
This is an extraordinary prediction, because the models are non-SUSY
overall. It is those superpartners that are responsible for
the breaking of the two additional $U(1)$ symmetries.
This is a very strong prediction since, given the fact that
the classes of models
are
non-SUSY, the string scale should be low in order for the hierarchy to be
stabilized.

\item Baryon number (B) is an unbroken gauged $U(1)$ symmetry, with the 
corresponding
gauge boson to receive a mass and the baryon number to survive as a global
symmetry to low energies. Thus proton
is stable.

\item Lepton number (L) is a gauged $U(1)$ symmetry, thus neutrinos 
have
Dirac masses. Lepton number remains as a global symmetry to low 
energies. The small masses for the neutrinos
will come
from the existence of chiral
condensates breaking the
PQ-like symmetry \footnote{The same PS like symmetry
was responsible
for giving small neutrino masses in the models of \cite{luis1, kokos2}.}
$U(1)_b$.
 The gauge bosons
corresponding to the gauging of B, L get a mass through their couplings to
RR fields that are being involved in a generalized
Green-Schwarz mechanism.

\end{itemize}

The paper is organized as follows.
In the next section we describe the rules
for constructing the six-stack models for the six-torus orientifolded,
with D6 branes intersecting at angles, constructions.
We present the representation content and the general solutions to
RR tadpole cancellation conditions for the classes of models giving 
rise exactly to the SM at low energies.
 In section 3 we describe
the cancellation of the mixed $U(1)$ gauge anomalies by a
dimensional reduction scheme which is equivalent to cancellation
of the field theory anomaly by its Green-Schwarz amplitude \cite{luis1}.
This mechanism has been 
used in the context of toroidal models with branes at angles
in \cite{allo3, luis1}. 
In section 4 we describe the electroweak Higgs sector of the models
giving our emphasis on the definition of the geometrical quantities
that characterize the geometry of the Higgs sector of the model.
In section 5 we describe the remarkable effect of
how by imposing the condition that
$N=1$ SUSY may be preserved by some
sectors, breaks the gauge symmetry to the SM itself.
In section 6, we present a case by case analysis of the possible
Higgs fields realized in the models as well describing the
neutrino mass generation.
Our conclusions together with some comments are presented in section 7.

\section{Exact Standard model compactifications from Intersecting branes}

The formalism that we will make use in this work 
is based on  
type I strings with D9-branes compactified on a six-dimensional
orientifolded torus $T^6$, 
where internal background 
gauge fluxes on the branes are turned on \cite{inter1, inter2}. 
By performing a T-duality transformation on the $x^4$, $x^5$, $x^6$, 
directions the D9-branes with fluxes are translated into D6-branes 
intersecting at 
angles. Also in this framework we note that the branes are not
paralled to the
orientifold planes.
Under the T-duality
the $\Omega $ symmetry,
where $\Omega$ is the worldvolume 
parity
is transformed into $\Omega {\cal R}$,
where $\cal R$ is the reflection on the T-dualized coordinates,
\beq
T (\Omega {\cal R})T^{-1}= \Omega {\cal R},
\label{dual}
\eeq
We assume that the D$6_a$-branes wrap 1-cycles 
$(n^i_a, m^i_a)$, $i=1,2,3$ along each of the ith-$T^2$
torus of the factorized $T^6$ torus, namely 
$T^6 = T^2 \times T^2 \times T^2$.
Thus we allow the six-torus to wrap factorized 3-cycles, so we can 
unwrap the 3-cycle into products of three 1-cycles, one for each $T^2$.
Defining the homology of the 3-cycles as
\beq
[\Pi_a] =\ \prod_{i=1}^3(n^i_a [a_i] + m^i_a[b_i])
\label{homo1}
\eeq
defines consequently the 3-cycle of the orientifold images as
\beq
[\Pi_{a^{\star}}] =\ \prod_{i=1}^3(n^i_a [a_i] - m^i_a[b_i])
\label{homo2}
\eeq

We note that in the presence of $\Omega {\cal R}$ symmetry, 
 each D$6_a$-brane
1-cycle, must be accompanied by its $\Omega {\cal R}$
orientifold image
partner $(n^i_a, -m^i_a)$.

The six-stack SM model structure
that we consider in this work is based on the stack structure
$U(3) \otimes U(2) \otimes U(1)_c \otimes U(1)_d \otimes U(1)_e
\otimes U(1)_f$ or
$SU(3) \otimes SU(2) \otimes U(1)_a \otimes U(1)_b \otimes U(1)_c \otimes 
U(1)_d \otimes U(1)_e \otimes U(1)_f $
at the string
scale.

In addition,  
the presence of discrete NS B-flux \cite{flux} is assumed. 
Thus, when the B-flux is present, the tori involved are not
orthogonal but tilted. In this way 
the wrapping numbers become the effective tilted wrapping numbers, 
\beq
(n^i, m ={\tilde m}^i + {n^i}/2);\; n,\;{\tilde m}\;\in\;Z.
\label{na2}
\eeq
Thus we allow semi-integer values for the m-wrapping numbers.

The chiral sector is computed from a number of different sectors.
As usual in these constructions the chiral fermions get localized in the
intersections between branes.
The possible sectors are \footnote{
We associate the action of 
$\Omega {\cal R}$ on a sector $a, b$, as being associated to its
images $a^{\star}, b^{\star}$, respectively.}:

\begin{itemize}
 
\item The $a b + b a$ sector: involves open strings stretching between the 
D$6_a$ and D$6_b$ branes. Under the $\Omega {\cal R}$ symmetry 
this sector is mapped to $a^{\star} b^{\star}
+ b^{\star} a^{\star}$ sector.
The number, $I_{ab}$, of chiral fermions in this sector, 
transform in the bifundamental representation
$(N_a, {\bar N}_a)$ of $U(N_a) \times U(N_b)$, and reads
\beq
I_{ab} = [\Pi_a] \cdot [\Pi_b] = ( n_a^1 m_b^1 - m_a^1 n_b^1)( n_a^2 m_b^2 - m_a^2 n_b^2 )
(n_a^3 m_b^3 - m_a^3 n_b^3),
\label{ena3}
\eeq
 where $I_{ab}$
is the intersection number of the wrapped cycles. Note that
 we denote the chirality of
the fermions as being associated to
the sign of
 $I_{ab}$ intersection,
 where $I_{ab} > 0$ denotes left handed fermions.
Moreover, with negative multiplicity we denote the opposite chirality.

\item The $a b^{\star} + b^{\star} a$ sector :
It involves chiral fermions transforming into the $(N_a, N_b)$
representation with multiplicity given by
\beq
I_{ab^{\star}} =\  [\Pi_a] \cdot [\Pi_{b^{\star}}]     =\ -( n_a^1 m_b^1 + m_a^1 n_b^1)( n_a^2 m_b^2 + m_a^2 n_b^2 )
(n_a^3 m_b^3 + m_a^3 n_b^3).
\label{ena31}
\eeq
The $\Omega {\cal R}$ symmetry transforms this sector to itself.

\item the $a a^{\star}$ sector : under the $\Omega R$ symmetry it 
transforms to itself. In this sector the invariant intersections
will give 8$m_a^1 m_a^2 m_a^3$ fermions in the antisymmetric representation
and the non-invariant intersections that come in pairs provide us with
4$ m_a^1 m_a^2 m_a^3 (n_a^1 n_a^2 n_a^3 -1)$ additional 
fermions in the symmetric and 
antisymmetric representation of the $U(N_a)$ gauge group.
As it will be
explained later, these
sectors will be absent from our models.
\end{itemize}

Any vacuum derived from the previous intersection constraints
is subject in addition to constraints coming from RR tadpole cancellation 
conditions \cite{inter1, inter2}.
That demands cancellation of
D6-branes charges \footnote{Taken together with their
orientifold images $(n_a^i, - m_a^i)$  wrapping
on three cycles of homology
class $[\Pi_{\alpha^{\prime}}]$.}, wrapping on three cycles with
homology $[\Pi_a]$ and the O6-plane 7-form
charges wrapping on 3-cycles  with homology $[\Pi_{O_6}]$. 
Note that the RR tadpole cancellation conditions can be expressed
in terms of cancellations of RR charges in homology as
\beq
\sum_a N_a [\Pi_a]+\sum_{\alpha^{\star}} 
N_{\alpha^{\star}}[\Pi_{\alpha^{\star}}] - 32
[\Pi_{O_6}]=0.
\label{homology}
\eeq  
In explicit form, the RR tadpole conditions read
\beqa
\sum_a N_a n_a^1 n_a^2 n_a^3 =\ 16,\nonumber\\
\sum_a N_a m_a^1 m_a^2 n_a^3 =\ 0,\nonumber\\
\sum_a N_a m_a^1 n_a^2 m_a^3 =\ 0,\nonumber\\
\sum_a N_a n_a^1 m_a^2 m_a^3 =\ 0.
\label{na1}
\eeqa
That quarantees absense of non-abelian gauge anomalies.

The complete accommodation
of the fermion structure of the six-stack SM model can be seen
in table (\ref{spectrum8}).
Several comments are in order:

\begin{table}[htb] \footnotesize
\renewcommand{\arraystretch}{1.5}
\begin{center}
\begin{tabular}{|c|c|c|c|c|c|c|c|c|c|}
\hline
Matter Fields & & Intersection & $Q_a$ & $Q_b$ & $Q_c$ & $Q_d$ & $Q_e$&
$Q_f$ & Y
\\\hline
 $Q_L$ &  $(3, 2)$ & $I_{ab}=1$ & $1$ & $-1$ & $0$ & $0$ & $0$&
 $0$  & $1/6$ \\\hline
 $q_L$  &  $2(3, 2)$ & $I_{a b^{\ast}}=2$ &  
$1$ & $1$ & $0$ & $0$  & $0$ &$0$ & $1/6$  \\\hline
 $U_R$ & $3({\bar 3}, 1)$ & $I_{ac} = -3$ & 
$-1$ & $0$ & $1$ & $0$ & $0$ & $0$ & $-2/3$ \\\hline    
 $D_R$ &   $3({\bar 3}, 1)$  &  $I_{a c^{\ast}} = -3$ &  
$-1$ & $0$ & $-1$ & $0$ & $0$ &$0$ & $1/3$ \\\hline    
$L^1$ &   $(1, 2)$  &  $I_{bd} = -1$ &  
$0$ & $-1$ & $0$ & $1$ & $0$ &$0$ & $-1/2$  \\\hline    
$L^2$ &   $(1, 2)$  &  $I_{b e} = -1$ &  
$0$ & $-1$ & $0$ & $0$ & $1$ & $0$ & $-1/2$  \\\hline
$L^3$ &   $(1, 2)$  &  $I_{b f} = -1$ &  
$0$ & $-1$ & $0$ & $0$ & $0$ & $1$ & $-1/2$  \\\hline
$N_R^1$ &   $(1, 1)$  &  $I_{cd} = 1$ &  
$0$ & $0$ & $1$ & $-1$ & $0$ & $0$ & $0$  \\\hline    
$E_R^1$ &   $(1, 1)$  &  $I_{c d^{\ast}} = -1$ &  
$0$ & $0$ & $-1$ & $-1$ & $0$ & $0$ &  $1$   \\\hline
  $N_R^2$ &   $(1, 1)$  &  $I_{c e} = 1$ &  
$0$ & $0$ & $1$ & $0$ & $-1$ & $0$ &  $0$ \\\hline
$E_R^{2}$ &   $(1, 1)$  &  $I_{c e^{\ast}} = -1$ &  
$0$ & $0$ & $-1$ & $0$ & $-1$ & $0$ &  $1$   \\\hline
$N_R^3$ &   $(1, 1)$  &  $I_{c f} = 1$ &  
$0$ & $0$ & $1$ & $0$ & $0$ & $-1$ &  $0$ \\\hline
$E_R^3$ &   $(1, 1)$  &  $I_{c f^{\ast}} = -1$ &  
$0$ & $0$ & $-1$ & $0$ & $0$  &$-1$ & $1$ \\\hline
\hline
\end{tabular}
\end{center}
\caption{\small Low energy fermionic spectrum of the six stack 
string scale 
$SU(3)_C \otimes
SU(2)_L \otimes U(1)_a \otimes U(1)_b \otimes U(1)_c 
\otimes U(1)_d \otimes U(1)_e \otimes U(1)_f$, type I D6-brane model 
together with its
$U(1)$ charges. Note that at low energies only the SM gauge group 
$SU(3) \otimes SU(2)_L \otimes U(1)_Y$ survives.
\label{spectrum8}}
\end{table}

$\bullet$
The models accommodate various known low energy gauged symmetries. 
The latter can be expressed 
in terms of the $U(1)$ symmetries $Q_a$, $Q_b$, $Q_c$, $Q_d$, $Q_e$, 
$Q_f$. 
We find the following identifications
\beqa
{\bf B}aryon \ number & \rightarrow& Q_a = 3{\bf B}, \nonumber\\
{\bf L}epton \ number  &\rightarrow& {\bf L} = Q_d + Q_e+ Q_f, \nonumber\\ 
3 ({\bf B-L}) &\rightarrow&  Q_a - 3 Q_d - 3 Q_e -3 Q_f.  \nonumber\\
\label{diag}
\eeqa

Moreover, $Q_c = 2I_{3R}$, where 
$I_{3R}$ being the third component of weak isospin. Also,
$3 (B-L)$ and $Q_c$ are free of triangle anomalies. The $U(1)_b$ symmetry
plays the role of a Peccei-Quinn symmetry in the sence of having mixed
SU(3) anomalies. This symmetry
appears to be a general feature, of the model building 
based orientifolded six-torus constructions with D6 branes 
intersecting at angles, in the models based on 
the four- \cite{luis1} and five-stack SM's \cite{kokos2}.

$\bullet$ 
The study of 
Green-Schwarz mechanism will show us   
that Baryon and Lepton number are unbroken gauged   
symmetries and the corresponding gauge bosons are massive.
It
is important to notice that baryon and lepton numbers remain as global
symmetries to low energies. 
Thus proton should be stable. Also Majorana masses for right handed
neutrinos are not allowed in the models, that is 
mass terms for neutrinos should be of Dirac type.  
In the SM only the diagonal combination
\beqa
{\bf L}_{diag} = L_e + L_{\mu}+ L_{\tau}
\label{exact}
\eeqa
 is an
exact symmetry, that means $L_{diag}$ is preserved
in each SM interaction. 
Thus it appears that the six stack SM's offer a very logical explanation
for the existence of the various global symmetries that
exist in the SM, in particular the fact the $U(1)_{B-L}$ is an exact 
global symmetry.
\newline  
$\bullet$ The mixed anomalies $A_{ij}$ of the six surplus $U(1)$'s 
with the non-abelian gauge groups $SU(N_a)$ of the theory
cancel through a generalized GS mechanism \cite{iru, sa},
involving
 close string modes couplings to worldsheet gauge fields.
 Two combinations of the $U(1)$'s are anomalous and become massive, their
 orthogonal 
 non-anomalous combinations survive, combining to a single $U(1)$
 that remains massless, the latter to be identified with the hypercharge
generator.
\newline
$\bullet$
The structure of intersection numbers which give the parametric form
of tadpole solutions is unique for a certain level of stacks of
branes. Another choise of intersection numbers at this level of stacks
neither produces the correct hypercharge assignments
for the SM chiral particles nor is able to produce a general class
of solutions like those presented here or in the four-, five- stack SM's
\cite{luis1, kokos2} respectively.
\newline
$\bullet$
The models make use of the constraint
\beq
 {\Pi}_{i=1}^3 m^i =\ 0. 
\label{req1}
\eeq
The latter constraint is essential to
cancel the appearance of exotic representations in the model,
appearing from sectors in the form ${\alpha} {\alpha}^{\star}$, in antisymmetric and 
symmetric 
representations of the $U(N_a)$ group.

$\bullet$ The solutions satisfying simultaneously the
intersection constraints and the
cancellation of the RR crosscap tadpole constraints
are given in parametric form in table (\ref{spectrum10}). 
These solutions represent
the most general solution of the RR tadpoles 
as they depend 
on six 
integer parameters $n_a^2$, $n_d^2$, $n_e^2$, $n_f^2$, $n_b^1$, $n_c^1$,
the phase parameter
$\epsilon = \pm 1$, the NS-background
parameter
$\beta_i =\ 1-b_i$, which is associated to the presence of the 
NS B-field by $b_i =0,\ 1/2$, and the interpolating parameter
${\tilde \epsilon} = \pm 1$ which gives two different classes of
RR tadpole solutions.

The multiparameter
tadpole solutions appearing in table (2) represent deformations of 
the D6-brane branes, of table (1), intersecting at angles,
 within the same homology class of the
factorizable three-cycles. 
The 
solutions of table (2) in (\ref{na1}) satisfy all tadpole equations but the 
first. The
\footnote{We have added an arbitrary number
of $N_D$ branes which don't contribute to the rest of the tadpoles and
intersection number constraints. Thus in terms of the low energy theory
they don't have no effect.}
latter becomes \footnote{We have set for simplicity
$\epsilon = {\tilde \epsilon}= 1$.} :

\begin{table}[htb]\footnotesize
\renewcommand{\arraystretch}{3}
\begin{center}
\begin{tabular}{||c||c|c|c||}
\hline
\hline
$N_i$ & $(n_i^1, m_i^1)$ & $(n_i^2, m_i^2)$ & $(n_i^3, m_i^3)$\\
\hline\hline
 $N_a=3$ & $(1/\beta^1, 0)$  &
$(n_a^2,  \epsilon \b^2)$ & $(3, {\tilde \epsilon}/2)$  \\
\hline
$N_b=2$  & $(n_b^1, -\epsilon \b^1)$ & $(1/\beta_2, 0)$ &
$({\tilde \epsilon}, 1/2)$ \\
\hline
$N_c=1$ & $(n_c^1, \epsilon \b^1)$ &   $(1/\beta^2, 0)$  & 
$(0, 1)$ \\    
\hline
$N_d=1$ & $(1/\beta^1, 0)$ &  $(n_d^2,   \epsilon \b^2)$  
  & $(1, -{\tilde \epsilon}/2)$  \\\hline
$N_e = 1$ & $(1/\beta^1, 0)$ &  $(n_e^2,   \epsilon \b^2)$  
  & $(1, -{\tilde \epsilon}/2)$  \\\hline
$N_f = 1$ & $(1/\beta^1, 0)$ &  $(n_f^2,   \epsilon \b^2)$  
  & $(1, -{\tilde \epsilon}/2)$  \\
\hline
\end{tabular}
\end{center}
\caption{\small Tadpole solutions for six-stacks of
D6-branes giving rise to, exactly, the
standard model gauge group and observable chiral spectrum,
at low energies.
The solutions depend 
on six integer parameters, 
$n_a^2$, $n_d^2$, $n_e^2$, $n_f^2$, $n_b^1$, $n_c^1$,
the NS-background $\beta^i$ and
the phase parameter $\epsilon = \pm 1$ and the extra
interpolating parameter
 ${\tilde \epsilon} = \pm 1$. The ${\tilde \epsilon}$
parameter distinguishes the two different classes of tadpole
solutions.
\label{spectrum10}}          
\end{table}

\beq
\frac{9 n_a^2}{ \b^1} + 2 \frac{n_b^1}{ \b^2} +
\frac{n_d^2}{ \b^1} + \frac{n_e^2}{ \b^1} + \frac{n_f^2}{ \b^1} 
+  N _D \frac{2}{\b^1 \b^2} =16.
\label{ena11}
\eeq
Note that we had added the presence of extra $N_D$ branes. 
Their contribution to the RR tadpole conditions is best 
described by placing them in the three-factorizable cycle 
\beq 
N_D (1/\b^1, 0)(1/\b^2, 0)(2, m_D^3)
\label{sda12}
\eeq
where we have set $ m_D^3 =0$. 
The cancellation of tadpoles is better seen, if we choose a
numerical set of wrappings, e.g.
\beq
n_a^2 =1,\;n_b^1 =1,\;n_c^1 \in Z,\;n_d^2 =-1,
\;n_e =-1,\;n_f^2=-1,\;\b^1 =1,\;\b^2 =1.
\label{numero1}
\eeq
Within the above choise, all tadpole conditions but the first are 
satisfied, the latter is satisfied when we add
four $D6$ branes, e.g. $N_D = 4$ positioned at $(1, 0)(1, 0)(2,0)$. 
Thus the tadpole structure \footnote{Note that the parameter
$n_c^1$ should be defined such that its choise is consistent with a 
tilted tori, e.g. $n_c^1 = 1$. } becomes
\begin{center}
\beqa
N_a =3&(1, \ 0)
(1, \  1)(3,  1/2) \nonumber\\
N_b =2&(1,  -1)(1, \ 0) (1, \ 1/2) \nonumber\\
N_c =1&(n_c^1, \ 1)(1,\ 0)(0,\ 1) \nonumber\\    
N_d =1&(1, \ 0)(-1, 1)(1, -1/2)\nonumber\\
N_e =1&(1,\ 0) (-1,\ 1)(1, \  -1/2)\nonumber\\
N_f =1&(1,\ 0) (-1,\ 1)(1, \  -1/2)  
\label{consist}
\eeqa
\end{center} 
Actually, the satisfaction of the tadpole conditions  
is independent of $n_c^1$. Thus, when all other parameters are fixed,
$n_c^1$ is a global parameter that can vary
according to if the first tori is, or not, tilted.
Its precise value will be
fixed in terms of the remaining tadpole parameters
when we determine the tadpole
subclass that is associated with the
 hypercharge
embedding of the standard model. \newline
Note that there are always choises of
of wrapping numbers that satisfy the RR tadpole 
constraints without the need of adding extra paralled branes, e.g. the
following choise satisfies all RR
tadpoles
\beqa
n_a^2 =1,\;n_b^1 =-1,\;n_c^1 \in 2Z+1,\;n_d^2 =0,\nonumber\\
n_e^2 =0,\ n_f^2 =0,\;\b^1 =1/2,\;\b^2 =1.
\label{numero2}
\eeqa 
with cycle wrapping numbers \footnote{Another consistent choise will 
be $\b^1 =1$, $\b^2= 1/2$, $n_d^2 = n_e^2 = n_f^2 = 1$, $n_a^2=1$, 
$n_b^1=1$.
We have set for simplicity $\epsilon =\ {\tilde  \epsilon} =\ 1$.  }
\begin{center}
\beqa
N_a =3&(2, \ 0)(1, \  1)(3,  1/2) \nonumber\\
N_b =2&(-1,  -1/2)(1, \ 0) (1, \ 1/2) \nonumber\\
N_c =1&(n_c^1, \ 1/2)(1,\ 0)(0,\ 1) \nonumber\\    
N_d =1&(2, \ 0)(0, 1)(1, \ -1/2)\nonumber\\
N_e =1&(2,\ 0) (0,\ 1)(1, \  -1/2)\nonumber\\
N_f =1&(2,\ 0) (0,\ 1)(1, \  -1/2)
\label{consist1}
\eeqa
\end{center}

$\bullet$ The hypercharge operator corresponding
to the spectrum of table (1),
is defined as a linear combination
of the $U(1)$ gauge groups, $U(1)_a$, $U(1)_c$, $U(1)_d$, $U(1)_e$, $U(1)_f$, as
\beq
Y = \frac{1}{6}U(1)_{a}- \frac{1}{2} U(1)_c - \frac{1}{2} U(1)_d -
 \frac{1}{2} U(1)_e -\frac{1}{2} U(1)_f\;.
\label{hyper12}
\eeq

\section{U(1) anomaly cancelation}

The general form 
the mixed anomalies $A_{ij}$ of the six $U(1)$'s
with the non-Abelian gauge groups are given by
\beq
A_{ij}= \frac{1}{2}(I_{ij} - I_{i{j^{\star}}})N_i.
\label{ena9}
\eeq
From the mixed anomalies 
of the $U(1)$'s with the non-abelian gauge groups $SU(3)_c$, 
$SU(2)_b$, we conclude that there are two anomaly free combinations
$Q_c$, $Q_a - 3 Q_d - 3 Q_e - 3 Q_f$.
Also the gravitational anomalies cancel since D6-branes never 
intersect O6-planes.
Gauge anomaly 
cancellation
\cite{iru}
in the orientifolded type I torus
models  is quaranteed through a 
generalized GS
mechanism \cite{allo3} that uses the 10-dimensional RR gauge fields
$C_2$ and $C_6$ and gives at four dimensions
the following couplings to gauge fields
 \beqa
N_a m_a^1 m_a^2 m_a^3 \int_{M_4} B_2^o \wedge F_a &;& n_b^1 n_b^2 n_b^3
 \int_{M_4}
C^o \wedge F_b\wedge F_b,\\
N_a  n^J n^K m^I \int_{M_4}B_2^I\wedge F_a&;&n_b^I m_b^J m_b^K \int_{M_4}
C^I \wedge F_b\wedge F_b\;,
\label{ena66}
\eeqa
where
$C_2\equiv B_2^o$ and $B_2^I \equiv \int_{(T^2)^J \times (T^2)^K} C_6 $
with $I=1, 2, 3$ and $I \neq J \neq  K $. We notice that 
the four dimensional duals
of $B_2^o,\ B_2^I$ are defined as:
\beqa
C^o \equiv \int_{(T^2)^1 \times (T^2)^2 \times (T^2)^3} C_6&;C^I \equiv
\int_{(T^2)^I} C_2, 
\label{ena7}
\eeqa
where $dC^o =-{\star} dB_2^o,\; dC^I=-{\star} dB_2^I$.

The cancellation of triangle anomalies (\ref{ena9}) derives 
from the existence of the
string amplitude involved in the GS mechanism \cite{sa} in four 
dimensions \cite{iru}. 
The latter amplitude, where the $U(1)_a$ gauge field couples to one
of the propagating
$B_2$ fields, that couples to dual scalars, and couples in turn to
two $SU(N)$ gauge bosons, is 
proportional \cite{luis1} to
\beq
-N_a  m^1_a m^2_a m^3_a n^1_b n^2_b n^3_b -
N_a \sum_I n^I_a n^J_a n^K_b m^I_a m^J_b m^K_b\; ,
I \neq J, K 
\label{ena8}
\eeq

Taking into account the constraint of 
(\ref{req1}) the RR couplings $B_2^I$ of (\ref{ena66}) then 
appear into the following three terms \footnote{For simplicity we have set
${\tilde \epsilon} =\ 1$.}:
\beqa
B_2^1 \wedge \left( \frac{- 2 \epsilon  \b^1 }{\b^2 } 
\right)F^b,&\nonumber\\         
B_2^2 \wedge \left(\frac{\epsilon \b^2}{\b^1}  
\right)(9F^a +  F^d+  F^e + F^f),&\nonumber\\
B_2^3  \wedge \left( \frac{3 {\tilde \epsilon} n_a^2}{2\b^1} F^a +     
\frac{n_b^1}{\b^2}F^b  + \frac{n_c^1}{\b^2} F^c -
\frac{{\tilde \epsilon} n_d^2}{2\b^1} F^d
-\frac{{\tilde \epsilon} n_e^2}{2 \b^1}F^e
-\frac{{\tilde \epsilon} n_f^2}{2 \b^1}F^f \right).&
\label{rr1}
\eeqa
Also
the couplings of the dual scalars $C^I$ of $B_2^I$ that are
required to cancel
the mixed anomalies of the six $U(1)$'s with the 
non-abelian gauge groups $SU(N_a)$ are given by
\beqa
C^1 \wedge [  \frac{\epsilon {\tilde \epsilon}\b^2 }{2 \b^1} (F^a \wedge 
F^a) - \frac{\epsilon {\tilde \epsilon} \b^2 }{ \b^1} (F^d \wedge F^d) -
\frac{ \epsilon  {\tilde \epsilon} \b^2 }{2 \b^1} F^e \wedge F^e
-
\frac{ \epsilon  {\tilde \epsilon} \b^2 }{2 \b^1} F^f \wedge F^f
)], &\nonumber\\
C^2 \wedge [ \frac{- \epsilon \b^1 }{2 \b^2 } (F^b \wedge 
F^b) +  \frac{ \epsilon \b^1 }{ \b^2 }(F^c \wedge F^c)    ],
&\nonumber\\
C^o \wedge [  \frac{3n_a^2}{\b^1}(F^a \wedge 
F^a)  +     \frac{{\tilde \epsilon}n_b^1}{ \b^2}(F^b \wedge 
F^b)  + \frac{n_d^2}{ \b^1}(F^d \wedge 
F^d)  + &\nonumber\\
+ \frac{n_e^2}{ \b^1}(F^e \wedge F^e 
+ \frac{n_f^2}{ \b^1}(F^f \wedge F^e) ].&
\label{rr2}
\eeqa

As in the four stack SM \cite{luis1}, or the five stack SM \cite{kokos2},
the RR scalar $B_2^0$ does not couple to any 
field $F^i$, as we have imposed the condition (\ref{req1}) 
which excludes the appearance of any exotic 
matter representations in the models.\newline
Note that these representations are not necessary in the SM based stack
constructions. However,
in the context of building a
GUT brane model that eventually has to break to the
SM they are welcome, as they become the reason for achieving the breaking
to the SM \cite{kokos3}.

A closer look at (\ref{rr1}) reveals that 
there are two anomalous $U(1)$'s that become massive through their 
couplings to the RR fields. They are the model independent fields,
 $U(1)_b $ and the combination $9U(1)_a +
 U(1)_d + U(1)_e +U(1)_f$,
which become massive through their couplings to the 
RR 2-form fields $B_2^1, B_2^2$ respectively. Also there is 
a model dependent, non-anomalous and massive $U(1)$ field
coupled to $B_2^3$ RR field.
Thus the 
 two non-anomalous free combinations are 
$U(1)_c$ and $U(1)_a - 3U(1)_d -3 U(1)_e -3 U(1)_f$.
In addition, we note that the 
 mixed anomalies $A_{ij}$ are cancelled 
by the GS mechanism set by the couplings (\ref{rr1}, \ref{rr2}).
\newline
The question we have to ask now, is how we can,
from the general
class 
of models, associated with the generic SM's of tables (1) and (2),
we can separate the subclass of models associated with
just the SM hypercharge
assignment at low energies.
 
The generalized Green-Schwarz mechanism
that cancels the non-abelian anomalies of the $U(1)$'s to the 
non-abelian gauge fields involves couplings of closed RR string modes 
to the $U(1)$ field strengths \footnote{In addition,
to the couplings of the Poincare dual scalars $\eta_a$ of the fields $B_a$,
\beq
\sum_a g_a^k \ \eta_a \ tr(F^k \wedge F^k).  
\label{gree1}
\eeq
}
in the form
\beq
\sum_a \ f_a^i \  B_a\wedge tr(F_i).
\eeq
Finally, the mixture of couplings in the form
\beq
A^{i k} +  \sum_a \ f_a^i \ g_a^k = \ 0 
\eeq
cancels all non-abelian $U(1)$ gauge anomalies.       
That means, as was argued in \cite{luis1}, that if we want to
keep some $U(1)$ massless we have to
keep it decoupled from some closed string mode couplings that can make 
it massive, that is
 \beq
\sum_a \ ( \frac{1}{6} {\tilde f}_a^{\alpha} -\ \frac{1}{2}{\tilde f}^c_a
 -\
\frac{1}{2}{\tilde f}_a^{d} -\  \frac{1}{2}{\tilde f}_a^{e} -\  \frac{1}{2}{\tilde f}_a^f) = \ 0\ .
\eeq
In conclusion,
the combination of the $U(1)$'s which remains light at low 
energies is 
\beq
3( n_a^2 + n_d^2 +  n_3^2 + n_f^2) \neq 0,\;\ Q^l = n_c^1 (Q_a -3 Q_d -3 Q_e -3 Q_f)
-\frac{3\b^2 {\tilde \epsilon}( n_a^2 + n_d^2 + n_e^2 + n_f^2)}{2 \b_1} Q_c   .
\label{hyper}
\eeq
The subclass of tadpole solutions of (\ref{hyper}) having the SM hypercharge
assignment at low energies is exactly the one, where
the combination (\ref{hyper}) is proportional to
(\ref{hyper12}). That is  
\beq
n_c^1 = \ \frac{{\tilde \epsilon}\b_2}{2\b_1}( n_a^2 +\  n_d^2 +\  n_e^2 +\ n_f^2).
\label{mashyper}
\eeq
Summarizing, we have found that as long as (\ref{mashyper}) holds,
we can identify $Q^l$ as the hypercharge generator, which gives at the
chiral fermions of table (1) their correct SM hypecharge assignments.
Moreover, there are two extra anomaly free $U(1)$'s beyond the
hypercharge
combination, which read

\beqa
&Q^{(5)} =\ \frac{3{\tilde \epsilon}}{2 \b_1}(n_d^2 + n_e^2 + n_d^2) 
(Q_a -\ 3Q_d -\ 3Q_e -\ 3Q_f) + 28 n_c^1 Q_c &\nonumber\\
&Q^{(6)} =\ Q_e - Q_f &
\label{extra}
\eeqa
In the next section, we will break these
$U(1)$ symmetries
 by requiring the intersections where the right handed neutrino 
is localized, to preserve $N =\ 1$ supersymmetry.
A comment is in order. The $U(1)$ combinations
$Q_d -\ Q_e$, $Q_d -\ Q_f$ are anomaly free, that is we could have chosen
either of them to be the $Q^{(6)}$ generator. 
Then the only difference with the
present choise (\ref{extra}) would be a different constraint on the RR tadple 
cancellation conditions, once we will later impose $N=1$ on an intersection. 

Lets us summarize.
Up to this
point the gauge group content of the model includes, beyond $SU(3)
\otimes SU(2) \otimes U(1)_Y$, the additional $U(1)$ symmetries, 
$Q^{(5)}$, $Q^{(6)}$ under which some of the chiral SM particles of
table (1) gets charged.
The extra $U(1)$ symmetries will be broken by imposing
some open string sectors to respect some amount of SUSY. In the latter case
the immediate
effect on obtaining just the SM at low energies will be two additional
linear conditions on the RR tadpole solutions of table (2).
We note that when $n_c^1 = 0$, it is possible
to have massless in the low energy spectrum both the $U(1)$ generators,
$Q_c$, and the B-L generator $(1/3)(Q_a -3 Q_d -3 Q_e -3Q_f)$ as long as
$n_c^1 = 0$, $n_a^2 = -n_d^2-n_e^2-n_f^2$.

\section{Higgs mechanism on open string sectors}

The mechanism of electroweak symmetry breaking
at the string theory level between intersecting branes
is not well understood but it is
believed to take place
either by using open string tachyons \cite{carlo1, inter2, luis1, kokos2, kokos3}
between paralled branes
or using brane 
recombination \cite{allo6a}. As the nature of the
latter procedure is topological, it cannot be described using field
theoretical methods.
In this work, we will follow the former method and we leave the latter method for some future study.

\subsection{The angle structure}

In the previous sections we have detailed 
the appearance in the R-sector of open string excitations with
$I_{ab}$ massless chiral fermions
in the D-brane intersections that 
transform under the bifundamental representations $(N_a, {\bar N}_b)$.
However, in backgrounds with intersecting 
branes, besides the actual
presence of massless fermions at each intersection, 
we have evident the presence of an equal number of
 massive scalars (MS), in the NS-sector, in exactly the same representations 
as the massless fermions \cite{luis1}. The mass of the these MS
is of order of the string scale. In some cases,
it is possible that some of those MS may become 
tachyonic, triggering a potential that looks like the Higgs potential 
of the SM,
especially when their mass, that depends on the 
angles between the branes,
is such that is decreases the world volume of the 
3-cycles involved in the recombination process of joining the two
branes into a single one \cite{senn}.

The models that we are describing, are based on orientifolded six-tori
on type IIA
strings. In those configurations the bulk has ${\cal N} = 4$ SUSY.
Lets us now give some details about the open string sector of the models.
In order to describe the open string spectrum we introduce a four dimensional
twist \cite{allo3, luis1} 
vector $\upsilon_{\theta}$, whose I-th entry
is given by  $\vartheta_{ij}$, with $\vartheta_{ij}$ the angle
between the branes $i$ and $j$-branes. After GSO projection
the states are labeled by a four dimensional twisted vector $r +
\upsilon_{\theta}$, where $\sum_I r^I =$odd
and $r_I  \in {\bf Z}, {\bf Z} {\bf + \frac{1}{2}}$
for NS, R sectors respectively. The Lorentz quantum numbers
are denoted by the last entry.
The mass operator for the states reads:

\beq
{\alpha^\prime} M^2_{ij} =\ \frac{Y^2}{4 \pi^2 \alpha^{\prime}} +\
N_{bos}(\vartheta) +\ \frac{(r + \upsilon)^2}{2} -\ \frac{1}{2} +\
E_{ij},
\label{mass}
\eeq
where $E_{ij}$ the contribution to the mass operator from bosonic
oscillators, and $N_{osc}(\vartheta)$ their number operator, with
\beq
E_{ij} =\  \sum_I \frac{1}{2}|\vartheta_I|(1 - |\vartheta_I|),
\label{bos}
\eeq
and $Y$ measures the minimum distance between branes for minimum
winding states.  \newline
If we
represent the twisted vector $r + \upsilon$,  by $(\vartheta_1,\vartheta_2,
\vartheta_3, 0)$, in the NS open string sector, the 
lowest lying states are given  \footnote{
we assumed $0\leq\vartheta_i\leq 1$ .} by:
{\small \beqa
\begin{array}{cc}
{\rm \bf State} \quad & \quad {\bf Mass} \\
(-1+\vartheta_1,\vartheta_2,\vartheta_3,0) & \alpha' M^2 =
\frac 12(-\vartheta_1+\vartheta_2+\vartheta_3) \\
(\vartheta_1,-1+\vartheta_2,\vartheta_3,0) & \alpha' M^2 =
\frac 12(\vartheta_1-\vartheta_2+\vartheta_3) \\
(\vartheta_1,\vartheta_2,-1+\vartheta_3,0) & \alpha' M^2 =
\frac 12(\vartheta_1+\vartheta_2-\vartheta_3) \\
(-1+\vartheta_1,-1+\vartheta_2,-1+\vartheta_3,0) & \alpha' M^2
= 1-\frac 12(\vartheta_1+\vartheta_2+\vartheta_3)
\label{tachdsix}
\end{array}
\eeqa}
Also 
the angles at the thirteen different intersections can be expressed
in terms of the tadpole solutions parameters.
Let us define the angles :
\beqa
{\tilde \theta}_1 \   = \ \frac{1}{\pi} tan^{-1}
\frac{\b^1 R_2^{(1)}}{n_b^1 R_1^{(1)}},  \
\theta_2 \  =   \  \frac{1}{\pi}
tan^{-1}\frac{\beta^2 R_2^{(2)}}{n_a^2 R_1^{(2)}}, \
{\tilde \theta}_3 \  = \  \frac{1}{\pi} tan^{-1}\frac{R_2^{(3)}}
{6 R_1^{(3)}},
 \nonumber \\
{\theta_1} \   = 
\ \frac{1}{\pi} tan^{-1}\frac{ \beta^1 R_2^{(1)}}{n_c^1 R_1^{(1)}},\;\
{ \theta^{\prime}}_2  \ 
 = \ \frac{1}{\pi} tan^{-1} \frac{ \beta^2 R_2^{(2)}}{n_d^2 R_1^{(2)}},   \;\
{\theta_3 } \  = \ \frac{1}{\pi} tan^{-1}
\frac{R_2^{(3)}}{2 R_1^{(3)}},\nonumber\\
 \ {\tilde \theta}_2 \  =   \  \frac{1}{\pi} tan^{-1}
\frac{\beta^2 R_2^{(2)}}{n_e^2 R_1^{(2)}},    \ ;{\bar {\theta}}_2  \  = \
\frac{1}{\pi} tan^{-1}
\frac{R_2^{(2)}}{n_f^2 R_1^{(2)}},
\label{angulos}
\eeqa
where $R^{(i)}_{1,2}$ are the compactification radii
for the three $i=1,2,3$ tori, namely
projections 
of the radii 
 onto the $X^{(i)}_{1,2}$ directions when the NS flux B field,
$b^i$, is turned on and we have chosen for convenience $\epsilon =
{\tilde \epsilon} = \ 1$.

At each of the thirteen non-trivial intersections 
we have the 
presense of four states $t_i , i=1,\cdots, 4$, associated
to the states (\ref{tachdsix}).
 Hence we have a total of
fifty two different massive scalars, with lowest lying spectrum,
in the model \footnote{ In figure one,
we can see the D6 branes angle setup in the present models.} .

\begin{figure}
\begin{center}
\centering
\epsfysize=10cm
\leavevmode
\epsfbox{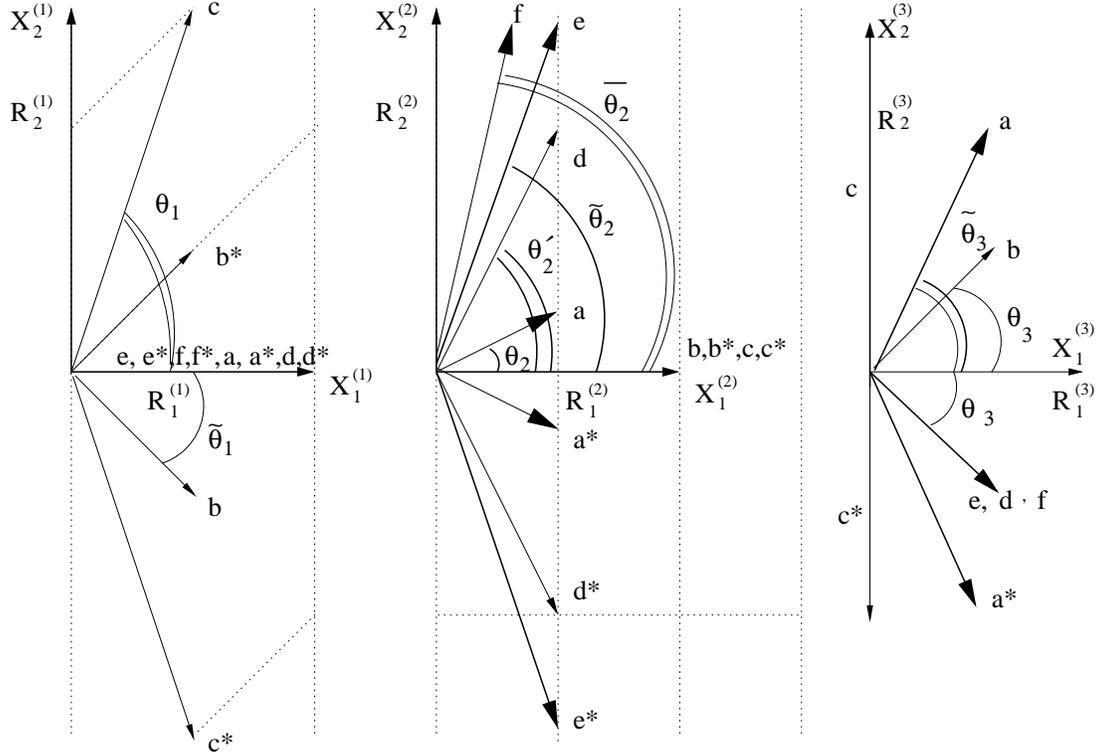}
\end{center}
\caption[]{\small
Assignment of angles between D6-branes on
the six stack type I model giving rise to the 
standard model at low energies.
The angles between branes are shown on a product of 
$T^2 \times T^2 \times T^2$. We have chosen $ \b^1 =\b^2 =1$, $n_b^1, 
n_c^1, n_a^2, 
n_d^2 >0$, $\epsilon = {\tilde \epsilon} = 1$. }
\end{figure}

The following mass relations hold between the different
intersections of the model :

\beqa
 m^2_{ab}(t_2)+m^2_{ac}(t_3) \ 
=\    m^2_{cd^{\star}}(t_2)+m^2_{cd^{\star}}(t_3)\
=\ m^2_{cd}(t_2)+m^2_{cd}(t_3)&\nonumber \\
= m^2_{ce}(t_1)+m^2_{ce}(t_3) \ 
=\    m^2_{ce^{\star}}(t_2)+m^2_{ce^{\star}}(t_3)\
=\ m^2_{cf}(t_2)+m^2_{cf}(t_3),
\nonumber \\
 m^2_{ab^{\star}}(t_2)+m^2_{ab^{\star}}(t_3) =
\ =\  m^2_{ab}(t_2)+m^2_{ab}(t_3)
\ =\  m^2_{bd}(t_2)+m^2_{bd}(t_3)&\nonumber \\
\ =\ m^2_{be}(t_2)+m^2_{be}(t_3) \ =\ m^2_{bf}(t_2)+
m^2_{bf}(t_3),&\nonumber \nonumber \\ 
 m^2_{be}(t_1)+m^2_{be}(t_2) \ = m^2_{bf}(t_1)+m^2_{bf}(t_2)
 \ = m^2_{bd}(t_1)+m^2_{bd}(t_2)\
\label{rela1}
\eeqa

We note that in this work, we will not discuss the stability conditions
for absence of tachyonic scalars such that the D-brane configurations
dicsussed will be stable as this will be discussed elsewhere.
Similar conditions have been examined before
in \cite{luis1, kokos2}.

\subsection{Tachyon Higgs mechanism}

In this section, we will study the electroweak Higgs sector
of the models.
We note that below the string scale the massless spectrum of
the model is that of the SM with all particles having the correct
hypercharge assignments
but with the gauge symmetry being
$SU(3) \otimes SU(2) \otimes U(1) \otimes Q^{(5)} \otimes Q^{(6)}$.
For the time being we will accept that the,
beyond the SM gauge group, $U(1)$ generators
break to a scale higher than
the scale of electroweak symmetry breaking. The latter issue will be
dicsussed in detail in the next section.

In general, tachyonic scalars stretching between two 
different branes  can be used as Higgs scalars
as they can become non-tachyonic
by varying the distance between paralled branes.
In the models presented the complex scalars
$h^{\pm}$, $H^{\pm}$ get
localized between the $b$, $c$ and between $b$, $c^{\ast}$ branes
respectively and 
can be interpreted from the field theory point of view \cite{luis1} 
as Higgs fields which are responsible for the breaking the
electroweak symmetry.
We note that the intersection numbers of the $b,c$ and $b,c^{\star}$
branes across the six-dimensional torus vanish as a result of the fact
that the $b, c$ and $b, c^{\star}$ branes are
paralled across the second tori.
The electroweak Higgs fields, appearing as
$H_i$ (resp. $h_i$), $i=1,2$, in table (\ref{hig}), come from the
NS sector, from open
strings stretching between the
paralled $b$, $c^{\star}$ (resp. $c$)
branes along the
second tori, and from open strings stretching between intersecting
branes along
the first and third tori.

Initially, the set of Higgs of table (\ref{hig}),
are part of the massive spectrum of fields
localized in the intersections $bc, bc^{\star}$. However, we note that
the Higges $H_i$, $h_i$ become massless by varying the distance along the
second tori between the $b, c^{\star}$, $b, c$ branes respectively.
Similar set of Higgs fields appear in the four stack
\cite{luis1} and five stack SM's \cite{kokos2} and the Pati-Salam four
stack models \cite{kokos3}, but obviously
with different geometrical data \footnote{All these models 
break exactly to the SM at low energies}.
We should note that the representations of Higgs fields $H_i$,  $h_i$ is the
maximum allowed by quantization. Their number is model dependent.

\begin{table} [htb] \footnotesize
\renewcommand{\arraystretch}{1}
\begin{center}
\begin{tabular}{||c|c||c|c|c||}
\hline
\hline
Intersection & EW breaking Higgs & $Q_b$ & $Q_c$ &Y \\
\hline\hline
$b c$ & $h_1$  & $1$ & $-1$&1/2\\
\hline
$b c$  & $h_2$  & $-1$ & $1$&-1/2 \\\hline\hline
$b c^{\star}$ & $H_1$  & $-1$ & $-1$& 1/2\\ \hline
$b c^{\star}$  & $H_2$  & $1$ & $1$ & -1/2\\
\hline
\hline
\end{tabular}
\end{center}
\caption{\small Higgs fields responsible for
electroweak symmetry breaking.
\label{hig}}
\end{table}

For the models presented in this work,
the number of complex scalar doublets
is equal to the non-zero
intersection number product between the $bc$, $bc^{\star}$
branes in the
first and third complex planes. Thus
\beqa
n_{H^{\pm}} =\  I_{bc^{\star}}\ =\ |\epsilon {\tilde \epsilon} \b_1(n_b^1 - n_c^1)|, &
n_{h^{\pm}} =\  I_{bc}\ =\ |\epsilon {\tilde \epsilon}\b_1(n_b^1 + n_c^1)| .
\label{inter1}
\eeqa

The precise geometrical data for the scalar doublets are 
{\small \beqa
\begin{array}{cc}
{\rm \bf State} \quad & \quad {\bf Mass^2} \\
(-1+\vartheta_1, 0, \vartheta_3, 0) & \alpha' {\rm (Mass)}_Y^2 =
  { {Z_2}\over {4\pi ^2}}\ +\ \frac{1}{2}(\vartheta_3 - 
\vartheta_1) \\
(\vartheta_1, 0, -1+ \vartheta_2, , 0) & \alpha' {\rm (Mass)}_X^2 =
  { {Z_3}\over {4\pi ^2 }}\ +\ \frac{1}{2}(\vartheta_1 - \vartheta_3) \\
\label{Higgsmasses}
\end{array}
\eeqa}
where  $X = \{ H_{bc^{\star}}^{+}, h_{bc}^{+}  \}$,
$Y = \{ H_{bc^{\star}}^{-}, h_{bc}^{-}  \}$   and  $Z_2$ is the
distance$^2$ in transverse space along the
second torus,
$\vartheta_1$, $\vartheta_3$ are the (relative) angles 
between the $b$-, $c^{\star}$ (for $H^{\pm}$ )  (or $b$, $c$ for $h^{\pm}$)
branes in the
first and third complex planes.  

As it have been discussed before in \cite{luis1, kokos2, kokos3, allo6} 
the presence of scalar doublets $H^{\pm}, h^{\pm}$, can be
seen as coming from the field theory mass matrix

\beq
(H_1^* \ H_2) 
\left(
\bf {M^2}
\right)
\left(
\begin{array}{c}
H_1 \\ H_2^*
\end{array}
\right)
+(h_1^* \ h_2)
\left(
\bf {m^2}
\right)
\left(
\begin{array}{c}
h_1 \\ h_2^*
\end{array}
\right) + h.c.
\eeq
where
\beqa
{\bf M^2}=M_s^2
\left(
\begin{array}{cc}
Z_2^{(bc^*)}&
\frac{1}{2}|\vartheta_1^{(bc^*)}-\vartheta_3^{(bc^*)}|  \\
\frac{1}{2}|\vartheta_1^{(bc^*)}-\vartheta_3^{(bc^*)}| &
Z_2^{(bc^*)}\\
\end{array}
\right),\\ \ {\bf m^2}=M_s^2
\left(
\begin{array}{cc}
Z_2^{(bc)}& \frac{1}{2}|\vartheta_1^{(bc)}-\vartheta_3^{(bc)}|  \\
\frac{1}{2}|\vartheta_1^{(bc)}-\vartheta_3^{(bc)}| & Z_2^{(bc)}
 \end{array} \right)
\eeqa
\vspace{1cm}
The fields $H_i$ and $h_i$ are thus defined as
\beq
H^{\pm}={1\over2}(H_1^*\pm H_2); \ h^{\pm}={1\over2}(h_1^*\pm h_2)  \ .
\label{presca}
\eeq

Hence the effective potential which 
corresponds to the spectrum of Higgs scalars is given by
\beqa
V_{Higgs}\ =\ m_H^2 (|H_1|^2+|H_2|^2)\
+\ m_h^2 (|h_1|^2+|h_2|^2)\  \nonumber\\
+\ m_B^2 H_1H_2\
+\ m_b^2 h_1h_2\
+\ h.c.,
\label{Higgspot}
\eeqa
where
\beqa
 {m_h}^2 \ =\ { {Z_2^{(bc)}}\over {4\pi ^2\alpha '}}\ & ; & \ 
{m_H}^2 \ =\ {{Z_2^{(bc^*)}}\over {4\pi ^2\alpha '}}\nonumber \\
m_b^2\ =\ \frac{1}{2\alpha '}|\vartheta_1^{(bc)}-\vartheta_3^{(bc)}| \ & ;&
m_B^2\ =\ \frac{1}{2\alpha '}|\vartheta_1^{(bc^*)}-\vartheta_3^{(bc^*)}|
\label{masilla}
\eeqa

We note that the $Z_2$ is a free parameter, a moduli,  
and can become very small 
in relation to the Planck scale.
However, the $m_B^2$, $m_b^2$ masses can be expressed in terms of
the scalar masses
of the particles present at the different intersections
\footnote{We have chosen a configuration
with $\epsilon = {\tilde \epsilon }= 1$,  $n_b, n_c, n_d, n_e > 0 $.}  :
\beqa
m_B^2\ =\ \frac{1}{2\alpha '}|-{\tilde \vartheta}_1 + \vartheta_1
 - \vartheta_3 -\frac{1}{2}| \ & ;&
m_b^2\ =\ \frac{1}{2\alpha '}|{\tilde \vartheta}_1 + \vartheta_1 
+ \vartheta_3
-\frac{1}{2}|\nonumber\\
m_h^2\ =\   \frac{1}{2\alpha '}(\chi_b -\chi_c)^2 &;& m_H^2\ =\ 
\frac{1}{2\alpha '}(\chi_b + \chi_c)^2,
\label{analy1} 
\eeqa 
where $\chi_b$, $\chi_c$ the distances from the orientifold plane of the 
branes b, c.
 Making use of the scalar mass relations at the intersections of the model
we can reexpress the mass relations (\ref{analy1}), in terms of 
(\ref{rela1}). Values of the $m_B^2$, $m_b^2$,
in terms of the scalar masses are given in Appendix I.

\section{Preserving $N=1$ SUSY at intersections }

Up to this point we have detailed the RR tadpole cancellation
solutions for the configuration of the chiral spectrum 
of table (1) to exist, as well deriving the hypercharge
condition (\ref{mashyper}) that quarantees that the hypercharge
survives massless to low energies.
However, we have not yet derived the gauge interactions
corresponding to the observable SM.
It remains to be proven
that the two additional
$U(1)$'s in (\ref{extra}) can be broken.
 The latter may happen by 
demanding that
some sector preserves $N=1$ SUSY. 
If this sector supports the presence of a scalar singlet
then we can, in principle, break the extra $U(1)$ generator
by giving a vev to the singlet.
Let us suppose that the sectors $ce$, $cf$ preserve $N=1$ SUSY.
In this case,  the immediate effect wil be the appearance
of $I_{ce}$, $I_{cf}$ massless scalars
at the intersections $ce$, $cf$ with the same quantum numbers
as the massless
$I_{ce}$, $I_{cf}$ fermions. Because $I_{ce}=1$, $I_{cf}=1$ and the massless
fermions localized
in the intersections $ce$, $cf$ respectively are $N_R^2$, $N_R^3$,
their massive partners, which will become
massless will be their superpartners, the ${\tilde N}_R^2$,
${\tilde N}_R^3$ singlet fields.

Lets us describe the procedure that we will follow in more detail.
In order for $N=1$ SUSY to be preserved at some intersection
between two branes $\gamma$, $\delta$ we need to satisfy
$ \pm \vartheta_1 \pm \vartheta_2 \pm \vartheta_3 = 0$ for some
choise of signs, where $\vartheta_i$, $i=1,2,3$ are the relative angles
of the branes $\gamma$, $\delta$ across the three 2-tori.

We want the particles localized on the intersection $ce$ to respect
some amount of SUSY, in our case $N =\ 1$.
 That means
that the relative angle between branes $c$, $e$, should obey the SUSY 
preserving condition
\beq    
\pm {\tilde \theta}_1 \pm {\bar \theta}_2 \pm (\frac{\pi}{2} + \theta_3)
=\ 0
\label{pre1}
\eeq

So lets us recast the SUSY condition for the $ce$ sector in the form
\beq
tan^{-1}{\frac{\b_1 U^{(1)}}{n_c^1}} + tan^{-1}\frac{\b_2 U^{(2)}}{n_e^2}
-  tan^{-1}(\frac{U^{(3)}}{2}) -\frac{\pi}{2} =\ 0, 
\label{sdad12}
\eeq
For the $cf$ sector the SUSY preserving condition  
\beq    
\pm {\theta}_1 \pm \theta^{\prime}_2 \pm (\frac{\pi}{2} + \theta^{\prime}_3)
=\ 0
\label{pre1}
\eeq
takes the form
\beq    
{\theta}_1 + \theta^{\prime}_2 - (\frac{\pi}{2} + \theta^{\prime}_3)
=\ 0
\label{pre2}
\eeq
Conditions  (\ref{sdad12}),  (\ref{pre2}) are solved by the choise
\beq
n_e^2 = 0,\ \; n_f^2 = 0,\\;\frac{\b_1 U^{(1)} }{n_c^1} 
=\ \frac{U^{(3)}}{2},\;\
U^{(i)} =\ \frac{R_2^{(i)} }{ R_1^{(i)}}.
\label{solu1}
\eeq
In particular the choise
\beq
n_e^2=0, \ n_f^2=0 
\label{condu}
\eeq
implies that the second tori is not
tilted, e.g. $ \b^2=1$.

For a set of Standard Model wrappings to exist we need to
consider both the
{\em hypercharge} (\ref{mashyper}) and the {\em gauge symmetry breaking}
conditions (\ref{solu1}). Thus when defining numerically the tadpole
solutions of table (2),
a consistent set for the observable SM
wrapping numbers will be
given by
\beq  
n_e^2 =0,\;n_f^2=0, \b^2=1,\;\b^1=1,\;n_b=-1,\;n_d^2=1,
\;n_a^2=1,\; n_c^1=1
\eeq
or
\begin{center}
\beqa
N_a=3&(1,  0)  (1,   1)  (3,   1/2) \nonumber\\
N_b=2&(-1,  -1)  (1,  0) (1,  1/2) \nonumber\\
N_c=1&(1, \  \ 1) (1, 0)  (0,\ \ 1)\nonumber\\    
N_d=1&(1,  0) (1, 1) (1, -1/2) \nonumber\\
N_e=1&(1, 0)  (0, 1)  (1, -1/2)\nonumber\\
N_f=1&(1, 0)  (0, 1)  (1, -1/2)
\eeqa
\end{center}
The latter choise satisfies all RR tadpole conditions but the first, 
the latter is satisfied
with the addition of three $D6$ branes located at 
$(1,0)(1,0)(2,0)$.

After imposing SUSY on sectors $ce$, $cf$, the
${\tilde N}_R^2$,
${\tilde N}_R^3$ scalar singlets
appear.
Consequently, as the singlets
${\tilde N}_R^2$,
${\tilde N}_R^3$
get vevs and get charged
under $Q^{(5)}$, $Q^{(6)}$ respectively, break the corresponding $U(1)$
gauge symmetries. Thus the final
symmetry leftover is exactly the
$SU(3) \otimes SU(2) \otimes U(1)_Y$, at low energies.
A consequence of the breaking is that this implies the existence of two
supersymmetric 
particles, massive 
supersymmetric partners of two left handed neutrinos, namely the 
${\tilde N}_R^2$, ${\tilde N}_R^3$.

\section{Neutrino couplings and masses}

In intersecting brane worlds the size of trilinear Yukawa
couplings e.g. between two lepton and a Higgs scalar is
controlled by the area of the worldsheet stretching among the
three D6-branes crossing at these intersections.
That means that the Yukawa couplings in our models are of order \cite{allo3}
\beq
Y_{ijk}= e^{-{\tilde A}_{ijk}}\  , 
\label{yuk}
\eeq
where ${\tilde A}_{klm}$ is the worldsheet area, in string units,
stretched between the three vertices
in the six dimensional compact space. The area of each
triangle stretching on each two-dimensional space can be expressed in terms
of the lengths of the sides a, b, c of the triangle using
the relation \footnote{
Found by \cite{heron}}, not widely
used at present,
\beq
A^{(2)} =\ \sqrt{s(s-a)(s-b)(s-c)},
\label{hronas}
\eeq
where $s=(a+b+c)/2$.

As we have said there are four different electroweak 
Higgs present in the models
$H_1$, $H_2$, $h_1$, $h_2$. Thus 
the full Yukawa interactions for the chiral spectrum of the SM's
fields, allowed by all the symmetries, read
\beqa
Y_j^U Q_L U_R^j h_1 +\ Y_j^D Q_L D_R^j H_2 &+\nonumber\\
Y_{ij}^u \ q_L^i \ U_R^j \ H_1 +\ Y_{ij}^d  \ q_L^i \ D_R^j \ h_2 &+\nonumber\\
\sum_{m=1}^3 Y^N_{ij} (L^m)^i (N_R^m)^j h_1 +\
\sum_{m=1}^3 Y_{ij}^E (L^m)^i (E_R^m)^j H_2 \ +\ h.c.& \nonumber\\
\label{era1} 
\eeqa
where $i=1, 2$, $j=1,2,3$.

\begin{table}
[htb] \footnotesize
\renewcommand{\arraystretch}{1.5}
\begin{center}
\begin{tabular}{||c|c|c|c|c|c|c||}
\hline\hline
Higgs fields & $\b_1$ & $\b_2$ & $n_a^2$ & $n_b^1$
& $n_c^1$ & $N_D$ \\\hline
$n_H=1,\  n_h =0$ & $1/2$ & $1$ & $-1-n_d^2$  & $1$ &$-1$ & $8 + 4n_d^2$ \\\hline
$n_H=1, \ n_h =0$ & $1/2$ & $1$ & $1-n_d^2$   & $-1$ & $1$ & $4n_d^2 $  \\\hline\hline
$n_H=0, \ n_h =1$ & $1/2$ & $1$ & $1-n_d^2$   & $1$ & $-1$ & $-1-4n_d^2$ \\\hline    
$n_h=0, \ n_h =1$ & $1/2$ & $1$ & $-1-n_d^2$  & $-1$ & $-1$ & $9+4n_d^2$ \\\hline\hline    
$n_H=1, \ n_h =1$ & $1$  &  $1$ &  $2-n_d^2$  & $0$ & $1$ & $-1-4 n_d^2$  \\\hline    
$n_H=1, \ n_h =1$ & $1$  &  $1$ &  $-2-n_d^2$ & $0$ & $-1$ & $17+4n_d^2$  \\\hline    
$n_H=1, \ n_h =1$ & $1$  &  $1$ &  $-n_d^2$   & $1$ & $0$ & $7 + 4n_d^2$   \\\hline
$n_H=1, \ n_h =1 $ & $1$  &  $1$ &  $-n_d^2$   & $-1$ & $0$ & $9+4n_d^2$ \\\hline    
\hline
\end{tabular}
\end{center}
\caption{\small Families of models with minimal Higgs structure. They
depend on a single integer, $n_d$. The surplus {\em gauge symmetry breaking}
condition (\ref{condu}) has been taken into account.
\label{tablo}}
\end{table}

The scalar doublets in the model present, that are
interpreted in terms of the low energy theory as Higgs doublets were
given in (\ref{presca}). As the number of Higgs present in the models 
depends on the parameters of the tadpole solutions, the most
interesting cases involve the following two possibilities:
The `minimal Higgs presence' case and the `next to minimal 
Higgs presence' case,
 $n_H = 1$, $n_h = 1$.

\begin{itemize}

\item {\em Minimal Higgs presence - Higgs system of MSSM}
\newline
In this case, we have
either $n_H = 1$, $n_h = 0$ or $n_H = 0$, $n_h = 1$. In those cases
we can see, see table (\ref{tablo}), that there are two families
of models left that depend on a single integer, e.g. $n_d^2$.
The solutions in this case are shown in the four top
rows of table (\ref{tablo}).
 We also list the number 
of necessary $N_D$ branes
 required to cancel the first tadpole condition.
 We have taken into account the conditions
 (\ref{mashyper}), (\ref{condu})
 necessary to obtain the observable SM at low energies.
The geometrical data for the Higgs system e.g. we choose $n_H = 1$,
$n_h = 0$ read :
\beqa
m_H^2 =\ \frac{(\chi_b + \chi_c)^2}{\alpha^{\prime}} ;,\;&
m_B^2 =\  \frac{1}{2 \alpha^{\prime} } |\vartheta_1 -
{\tilde \vartheta}_1 -\vartheta_3 -\frac{1}{2}|
\label{geom}
\eeqa

 A general feature of the SM models
 with four \cite{luis1} of five stacks \cite{kokos3} is the
 realization of the two Higgs system like in the MSSM.
In this minimal case, only the $H_1$, $H_2$ Higgs fields appear.
For the quark sector the analysis is identical to the one appearing
in \cite{luis1, kokos2}. From (\ref{era1}) we conclude that all charged
lepton get a tree level mass. Also 
two u-quarks and one d-quark receive a mass, namely the t, c, b quarks,
with masses of order of GeV.
The rest of the quarks,  don't receive a mass at tree level, as 
effective couplings in the form $Q_L U_R H_1$,  $q_L D_R H_2$
could violate the $U(1)_b$ PQ-like global symmetry.
However, this symmetry  will be 
broken by strong interaction effects, effectively giving masses to   
u, d, s quarks that have masses less than the QCD scale.\newline
As far as the neutrino masses are concerned, 
the models admits the following dimension six operators
\beq
L^1 N_R^1 (Q_L U_R)^{*},\;L^2 N_R^2 (Q_R U_R)^{*},
\;L^3 N_R^3 (Q_L U_R)^{*},\; 
\label{R}
\eeq
For values of the u-quark chiral condensate taken from lattice
calculations \cite{pilar}, with $<u_L u_R> \approx (240 MeV)^3$, and
values of the
string scale between 
1-10 TeV, one can get neutrino values between 0.1-10 eV in consistency
with neutrino oscillation experiments.

\item {\em Next to minimal Higgs presence} \newline
The next to minimal set of Higges is obtained
when $n_H=1, \ n_h =1$. In this
case, quarks and leptons get their mass from the start. 
From table (\ref{tablo})
we can see that there are two families of models that
depend on a
single integer $n_d^2$, we have imposed
the conditions (\ref{mashyper})
and (\ref{condu}).
 The geometrical data for this Higgs system read :
\beqa
m_H^2 = \frac{(\chi_b + \chi_c)^2}{\alpha^{\prime}}, \ \
m_h^2 = \frac{(\chi_b - \chi_c)^2}{\alpha^{\prime}}&\nonumber\\
m_B^2 =\  \frac{1}{2 \alpha^{\prime} } |\vartheta_1 - 
\vartheta_3 |,\ m_b^2 =\  \frac{1}{2 \alpha^{\prime} } |\vartheta_1 + 
\vartheta_3 |
\label{geom}
\eeqa

 A hierachy of lepton masses in this case would arise from a 
hierarchy of the Yukawa couplings and the vevs of Higgs masses involved.

\end{itemize}

\section{Conclusions and future directions}

In this paper, we have discuss, the construction and the
phenomenological properties of a new vacuum of type IIA theory 
compactified
on a orientifolded T6 torus with D6 branes intersecting an angles.
This particular construction gives us classes of models that have at low 
energies exactly the standard model. We note that until recently, 
obtaining the standard model at low energies from a
particular string construction
was an impossible task.

The models have some important properties 
including a stable proton, small neutrino masses.
Moreover, even if the models have a non-supersymmetric
spectrum they predict
the existence of exactly two supersymmetric 
particles, superpartners of two generations right handed
neutrinos. This phenomenon, namely the existence
of supersymmetric particles in a non-SUSY model is a completely new
phenomenon and has appeared before only in the five stack models we 
discussed in \cite{kokos3}. We note that the existence
of SUSY particles,
does not appear in the four stack
standard models of \cite{luis1} but it is a feature of its higher stack
generalizations \footnote{See also \cite{kokos2}.}. The reason that
supersymmetry make its appearance is
because of the need to create 
extra singlets into the theory that would break the
extra $U(1)$ generators present at low energies.

We also note that it is not possible in the present 
context \footnote{This has been confirmed in the cases
of n=5 \cite{kokos2} and n=6 in the present paper.} of 
stacks of D6 branes based on an $SU(3) \otimes SU(2) \otimes U(1)^n$
content \footnote{moving in a orientifolded six torus
background of IIA theory},
to lower the string scale with any known procedure.
This should be contrasted with the classes of models
discussed in \cite{kokos3}, 
where we have stacks of D6 branes, in the same compact background,
but with stacks of D6 branes based on a GUT like group
like $SU(4)_C  \otimes SU(2)_L \otimes SU(2)_R \otimes U(1)^4 $. 
In the latter case the string scale can be shown to be very low 
and constrained to be less than
650 GeV. The models presented, as well all the models derived from the 
same six torus backgrounds, have non-vanishing NSNS tadpoles.
Thus the question
 of the full
stability \footnote{For some other works
related to stability questions  on intersecting brane backgrounds
see also \cite{luis1, kokos2, blume, raba, blume1}. }
 of the models is an open question. It will be interesting
to examine numerically the question of full stability of our 
configurations along the 
lines of \cite{blume, blume1, raba}.
 We remind
that the background could be cured in principle
be redefining it \cite{fi, nsns} or by modifying the models such that
 the NSNS tadpoles may be absent. We hope to return to the last
issue in a future work.
Moreover, it will be interesting to examine the consequences for the 
electroweak breaking
for the present models and for the 5-stack SM's 
\cite{kokos2}
using brane recombination \cite{allo6aa}. 
We leave this task for a future work.\newline
Recently, it has been pointed out \cite{allo8} that it is possible,
by considering intersecting branes with 
compactifications of IIA theory on Calabi-Yau 3-folds, 
to rederive the chiral context of the 4-stack SM configurations 
of \cite{luis1}. It will be interesting to examine how the classes of
6-stack chiral 
configurations examined in the present work and those of 
5-stack SM's \cite{kokos2} can be 
realized explicitly in the Calabi-Yau case.

\section{Acknowledgements}
I am grateful to L. Ib\'a\~nez, R. N. Mohapatra,
S. Sint and A. Uranga for usuful discussions.

\section{Appendix A}

In this appendix, we list the values of the Higgs mass parameters, of section
(4).  Their values can be expressed in terms of the scalar masses at the 
different intersections.  

\begin{center}
\beqa
m_B^2 =\ |-\frac{1}{2}( m^2_{q_L }(t_2) + m^2_{q_L }(t_3)) + 
\frac{1}{2}( m^2_{U_R }(t_2) + m^2_{U_R }(t_3)  )-\nonumber\\
\frac{1}{4}(
 m^2_{L^2 }(t_1) + m^2_{L^2 }(t_2) ) -
\frac{1}{2}|=\nonumber\\
 |-\frac{1}{2}( m^2_{Q_L }(t_2) + m^2_{Q_L }(t_3)) + 
\frac{1}{2}( m^2_{N_R^2}(t_1) + m^2_{N_R^2 }(t_3)  )-\nonumber\\
\frac{1}{4}(
 m^2_{L^3 }(t_1) + m^2_{L^3 }(t_2) ) -
\frac{1}{2}|=\nonumber\\
 |-\frac{1}{2}( m^2_{L^1 }(t_2) + m^2_{L^1 }(t_3)) + 
\frac{1}{2}( m^2_{N_R^3 }(t_2) + m^2_{N_R^3 }(t_3)  )-\nonumber\\
\frac{1}{4}(
 m^2_{L^1 }(t_1) + m^2_{L^1 }(t_2) ) -
\frac{1}{2}|=\nonumber\\
 |-\frac{1}{2}( m^2_{L^3 }(t_2) + m^2_{L^3 }(t_3)) + 
\frac{1}{2}( m^2_{N_R^3 }(t_2) + m^2_{N_R^3 }(t_3)  )-\nonumber\\
\frac{1}{4}(
 m^2_{L^3 }(t_1) + m^2_{L^3 }(t_2) ) -
\frac{1}{2}|=\nonumber\\
 |-\frac{1}{2}( m^2_{E_R^1 }(t_2) + m^2_{E_R^1 }(t_3)) + 
\frac{1}{2}( m^2_{L^2 }(t_2) + m^2_{L^2}(t_3)  )-\nonumber\\
\frac{1}{4}(
 m^2_{L^1 }(t_1) + m^2_{L^1 }(t_2) ) -
\frac{1}{2}|\nonumber\\
\eeqa
\end{center}
\begin{center}
\beqa
m_b^2 =\ |\frac{1}{2}( m^2_{q_L }(t_2) + m^2_{q_L }(t_3)) + 
\frac{1}{2}( m^2_{N_R^2 }(t_1) + m^2_{N_R^2 }(t_3)  )+\nonumber\\
\frac{1}{4}(
 m^2_{L^1 }(t_1) + m^2_{L^1 }(t_2) ) -
\frac{1}{2}|=\nonumber\\
|\frac{1}{2}( m^2_{U_R }(t_2) + m^2_{U_R }(t_3)) + 
\frac{1}{2}( m^2_{Q_L }(t_2) + m^2_{Q_L }(t_3)  )-\nonumber\\
\frac{1}{4}(
 m^2_{L^3 }(t_1) + m^2_{L^3 }(t_2) ) -
\frac{1}{2}|=\nonumber\\
|\frac{1}{2}( m^2_{L^2 }(t_2) + m^2_{L^2 }(t_3)) + 
\frac{1}{2}( m^2_{E_R^2 }(t_2) + m^2_{E_R^2 }(t_3)  )+\nonumber\\
\frac{1}{4}(
 m^2_{L^2 }(t_1) + m^2_{L^2 }(t_2) ) -
\frac{1}{2}|=\nonumber\\
|\frac{1}{2}( m^2_{E_R^1 }(t_2) + m^2_{E_R^1 }(t_3)) + 
\frac{1}{2}( m^2_{L^3 }(t_2) + m^2_{L^3 }(t_3)  )+\nonumber\\
\frac{1}{4}(
 m^2_{L^2 }(t_1) + m^2_{L^2 }(t_2) ) -
\frac{1}{2}|\nonumber\\
\eeqa
\end{center}


\newpage

\begin{thebibliography}{99}
\small 

\bibitem{see} For reviews on string phenomenology
see (and references there in):\\
L.~E.~Ib\'a\~nez, hep-ph/9911499; \\ 
F. Quevedo, hep-ph/9707434; hep-th/9603074; \\
I. Antoniadis, hep-th/0102202, E. Dudas, hep-ph/0006190;\\
K. Dienes, hep-ph/0004129;hep-th/9602045


\bibitem{inter1}R.~Blumenhagen, L.~G\"orlich, B.~K\"ors and D.~L\"ust,
``Noncommutative compactifications of type I strings on tori with magnetic
background flux'',
JHEP 0010 (2000) 006, {\tt hep-th/0007024};
``Magnetic Flux in Toroidal Type I Compactification'', Fortsch. Phys. 49
(2001) 591, hep-th/0010198 

\bibitem{inter2}R. Blumenhagen, B. K\"ors and D. L\"ust,
``Type I Strings
with F and B-flux'',
JHEP 0102 (2001) 030, hep-th/0012156.

\bibitem{ellisnano}J. Ellis, S. Kelley, D. V. Nanopoulos, Phys. Lett. B249 
(1990) 441; U. Amaldi, W. De Boer 
and H. F\"urstenau, Phys. Lett. B260 
(1991) 131; P. Langacker and M. Luo, Phys. Rev. D44 (1991) 817 




\bibitem{kokos1}L. Dixon, V. Kaplunovsky and J. Louis,
Nucl Phys. B355 (1991) 649;\\
C. Kokorelis, 
``String Loop Threshold Corrections for N=1 Generalized Coxeter Orbifolds'',
Nucl.Phys. B579 (2000) 267-274, hep-th/0001217 ;\\
 D. Bailin, A. Love, W.A. Sabra, S. Thomas, 
``String Loop threshold corrections  for ${ Z}_N$ Coxeter orbifolds'',
Mod. Phys. Lett. A9 (1994) 67-80, hep-th/9310008; \\
C. Kokorelis, ``Gauge and Gravitational Couplings from Modular Orbits
in Orbifold
Compactifications'', Phys. Lett. B477 (2000) 313, hep-th/0001062\\
G. L. Cardoso, D. L\"ust, T. Mohaupt,  
``Threshold Corrections and Symmetry Enhancement in String Compactifications''
Nucl.Phys. B450 (1995) 115, hep-th/9412209



\bibitem{anto}I. Antoniadis, N. Arkadi-Hamed, S. Dimopoulos,
G. Dvali, Phys. Lett. B436
(1998) 257, hep-ph/9804398;
I. Antoniadis, C. Bachas, Phys. Lett. B450 (1999) 83


\bibitem{ber}M. Berkooz, M. R. Douglas, R.G. Leigh, 
``Branes Intersecting at Angles'',
\NPB480 (1996) 265, 
hep-th/9606139



\bibitem{flux}
M.~Bianchi, G.~Pradisi and A.~Sagnotti,
``Toroidal compactification and symmetry breaking in open string theories,''
Nucl.\ Phys.\ B376, 365 (1992);
Z.~Kakushadze, G.~Shiu and S.-H.~H.~Tye,
``Type IIB orientifolds with NS-NS antisymmetric tensor backgrounds,''
Phys.\ Rev.\ D58, 086001 (1998).
hep-th/9803141;
C. Angelantonj, \NPB 566 (2000) 126, 
``Comments on Open-String Orbifolds with a Non-Vanishing 
$B_{ab}$'', hep-th/9908064


\bibitem{carlo1}
R. Blumenhagen, L. G\"orlish, and B. K\"ors, ``Asymmetric Orbifolds, 
non-commutative geometry and type I string vacua'', 
Nucl. Phys. B582 (2000) 44, hep-th/0003024

\bibitem{carlo2}
C. Angelantonj and A. Sagnotti, 
``Type I vacua and brane transmutation'', hep-th/00010279;\\
C. Angelantonj, I. Antoniadis, E. Dudas and A. Sagnotti,
``Type I strings on magnetized orbifolds and brane transmutation'',
Phys. Lett. B489 (2000) 223, hep-th/0007090



\bibitem{luis1}
L.~E.~Ib\'a\~nez, F.~Marchesano and R.~Rabad\'an,
``Getting just the Standard Model at Intersecting Branes''
JHEP, 0111 (2001) 002, {\tt hep-th/0105155};

\bibitem{kokos2}C. Kokorelis, ``New Standard Model Vacua from
Intersecting Branes'', hep-th/0205147


\bibitem{kokos3}C. Kokorelis, ``GUT Model Hierarchies from Intersecting 
Branes'', hep-th/0203187



\bibitem{allo2}R. Blumenhagen, L. G\"orlich, B. K\"ors,
``Supersymmetric Orientifolds in 6D with D-Branes at Angles'',
Nucl.Phys. B569 (2000) 209,
hep-th/9908130; 
R. Blumenhagen, L. G\"orlich, B. K\"ors,
``Supersymmetric 4D Orientifolds of Type IIA with D6-branes at
Angles'',
JHEP 0001 (2000) 040,hep-th/9912204;
R. Blumenhagen, B. K\"ors, D. L\"ust and T. Ott,
``The Standard Model from Stable Intersecting Brane World Orbifolds'',
Nucl. Phys. B616 (2001) 3, hep-th/0107138

\bibitem{allo3}
G.~Aldazabal, S.~Franco, L.~E.~Ib\'a\~nez, R.~Rabad\'an and
A.~M.~Uranga,
``D=4 chiral string compactifications from intersecting branes'',
J. Math. Phys. 42 (2001) 3103-3126,
{\tt hep-th/0011073};
G.~Aldazabal, S.~Franco, L.~E.~Ib\'a\~nez, R.~Rabad\'an and
A.~M.~Uranga,
``Intersecting brane worlds'',
JHEP 0102 (2001) 047, {\tt hep-ph/0011132}.




\bibitem{allo4}Stefan Forste, Gabriele Honecker, Ralph Schreyer,
``Supersymmetric $Z_N \times Z_M$ Orientifolds in 4D with
D-Branes at Angles'',
Nucl.Phys. B593 (2001) 127-154, hep-th/0008250;
Ion V. Vancea,
``Note on Four Dp-Branes at Angles'',
JHEP 0104:020,2001,
hep-th/0011251;

\bibitem{allo5}H. Kataoka, M. Shimojo, `` $SU(3) \times SU(2) \times
U(1)$
Chiral models from Intersecting D4-/D5-branes'', hep-th/0112247;
G. Honecher, ``Non-supersymmetric Orientifolds with
D-Branes at Angles'', hep-th/0112174
G. Honecher, ``Intersecting brane world models from D8-branes on
$(T^2 \times T^4/Z_3)/\Omega {\cal R}_1$ type IIA orientifolds'',
hep-th/0201037;


\bibitem{allo6}D. Cremades,  L.~E.~Ib\'a\~nez, F. Marchesano,
`SUSY Quivers, Intersecting Branes and the Modest Hierarchy Problem',
hep-th/0201205;
D. Cremades, L. E. Ibanez and F. Marchesano,
 ``Standard Model at Intersecting D4/D5 Branes: Lowering the
String Scale'', hep-th/0205074;

\bibitem{allo6aa}D. Cremades, L. E. Ibanez and F. Marchesano,
`` Intersecting Brane Models of Particle Physics and the Higgs Mechanism'',
hep-th/0203160;


\bibitem{allo6a} L. F. Alday and G. Aldazabal, ``In quest of 
"just" the Standard Model on D-branes at a singularity'', 
JHEP 0205 (2002) 022, hep-th/0203129




\bibitem{allo7}G Altazabal, L.~E.~Ib\'a\~nez, A. M  Uranga,
`` Gauging Away the String CP problem'', hep-ph/0205250

\bibitem{allo8}R. Blumenhagen, V. Braun, B. K\"ors, D. L\"ust
`` Orientifolds of K3 and Calabi-Yau Manifolds with Intersecting D-branes'',
hep-th/0206038






\bibitem{shiu}M. Cvetic, G. Shiu A M. Uranga, ``Chiral four
dimensional $N=1$ supersymmetric IIA orientifolds from intersecting
D6 branes'', Nucl. Phys. B615 (2001) 3, hep-th/0107166;
M. Cvetic, G. Shiu A M. Uranga,
``Three family supersymmetric standard models from
intersecting brane worlds'', Phys. Rev. Lett. 87 (2001) 201801,
hep-th/0107143; \\
M. Cvetic, G. Shiu A M. Uranga, ``Chiral type II orientifold 
constructions as M theory on $G_2$ holonomy
  spaces'', hep-th/0111179; 
M. Cvetic, P. Laugacker, G. Shiu, ``Phenomenology
of a Three family Standard-like String Model '', hep-ph/0205252;
 M. Cvetic, P. Langacker, G. Shiu, 
``A Three-Family Standard-like Orientifold Model: 
Yukawa Couplings and Hierarchy'', hep-th/0206115
    
       







\bibitem{blume}R.~Blumenhagen, B.~K\"ors, D.~L\"ust and T. Ott,
``The Standard Model from Stable Intersecting Brane World Orbifolds'',
\NPB616 (2001) 3, hep-th/0107138





\bibitem{sa}A. Sagnotti,
``A Note on the Green - Schwarz Mechanism in Open - String Theories'',
 Phys. Lett. B294 (1992) 196, hep-th/9210127


\bibitem{iru}L.~E.~Ib\'a\~nez, R.~Rabad\'an and A. M. Uranga,
``Anomalous U(1)'s in Type I and Type IIB D=4, N=1 string vacua'',
Nucl.Phys. B542 (1999) 112-138




\bibitem{senn}A. Sen, JHEP 9808 (1998) 012, 
``Tachyon Condensation on the Brane Antibrane System''
hep-th/9805170; 
``SO(32) Spinors of Type I and Other Solitons on Brane-Antibrane
Pair'', JHEP 809 (1998) 023,
hep-th/9808141.




\bibitem{pilar}P. Hernandez, K. Jansen, L. Lellouch, H. Wittig,
``Scalar condensate and light quark masses from overlap fermions'', 
Nucl. Phys. Proc. Suppl. 106 (2002) 766, hep-lat/0110199


\bibitem{heron} Heron of Alexandria, ``Metrica'', Book I, 50 A.D  

\bibitem{blume1}R.~Blumenhagen, B.~K\"ors, D.~L\"ust and T. Ott,
``Moduli Stabilization for Intersecting Brane Worlds in type 0' String
Theory'', hep-th/0202124


\bibitem{raba}J. Garcia-Bellido and R. Rabadan, ``Complex Structure
Moduli Stability in Toroidal Compactifications'', JHEP 0205 (2002) 042,
hep-th/0203247






\bibitem{fi}W. Fischer and L. Susskind, Phys. Lett. B171 (1986) 383;
Phys. Lett. B173 (1986) 262

\bibitem{nsns}
E.~Dudas and J.~Mourad,
``Brane solutions in strings with broken supersymmetry and dilaton
 tadpoles'',
Phys. Lett. B486 (2000) 172, hep-th/0004165; R.~Blumenhagen, A.~Font,
``Dilaton tadpoles, warped geometries and large extra dimensions for
nonsupersymmetric strings'',
Nucl. Phys. B 599 (2001) 241, {\tt hep-th/0011269}.







\end{thebibliography}
\end{document}